\documentclass[%
 rsi,
amsmath,amssymb,
reprint,%
]{revtex4-1}

\usepackage{graphicx}% Include figure files
\usepackage{dcolumn}% Align table columns on decimal point
\usepackage{bm}% bold math
\usepackage[utf8]{inputenc}
\usepackage[T1]{fontenc}
\usepackage{mathptmx}
\usepackage{etoolbox}
\usepackage{epstopdf}
\usepackage{braket}
\usepackage{mathrsfs}
\usepackage{amsmath}
\usepackage{longtable}

\newcommand{\cost}{\$8000}

\makeatletter
\def\@email#1#2{%
 \endgroup
 \patchcmd{\titleblock@produce}
  {\frontmatter@RRAPformat}
  {\frontmatter@RRAPformat{\produce@RRAP{*#1\href{mailto:#2}{#2}}}\frontmatter@RRAPformat}
  {}{}
}%
\makeatother
\begin{document}

\title{An Inexpensive, Configurable Two-Tone Electron Spin Resonance Spectrometer}

\author{Charles A. Collett}
\affiliation{Department of Physics, Hamilton College, Clinton, NY 13323, USA}
\email{cacollet@hamilton.edu}
\author{Sofia M. Davvetas}
\affiliation{Department of Physics, Muhlenberg College, Allentown, PA 18104, USA}
\author{Abdulelah Alsuhaymi}
\author{Grigore A. Timco}
\affiliation{School of Chemistry, The University of Manchester, Manchester M13 9PL, UK}
\date{\today}
\begin{abstract}
Electron spin resonance (ESR) is a powerful tool for characterizing and manipulating spin systems, but commercial ESR spectrometers can be expensive and designed to work in narrow frequency bands. This work presents an inexpensive spectrometer that, when coupled with easy-to-design resonators, enables ESR over a broad frequency range, including at frequencies outside the standard bands. It can operate at either a single frequency or at two frequencies simultaneously. The spectrometer is built from off-the-shelf parts and controlled by a field programmable gate array (FPGA), and new capabilities can be easily added by reconfiguring the FPGA and adding or swapping components. We demonstrate the capabilities of the spectrometer using the molecular nanomagnet Cr$_7$Mn.%, and provide all the relevant control and acquisition code that we used.
\end{abstract}
\maketitle

\section{Introduction}
%INFO: RSI Structure: Title, author(s), affiliation(s), abstract, text, conclusion, supplementary material section, acknowledgments, author declarations section (conflict of interest, ethics approval, and author contributions), data availability statement, appendixes (if any), and references.

Electron spin systems are a subject of significant interest for the possibility of forming spin qubits \cite{nielsen_quantum_2010,loss_quantum_1998,chiesa_molecular_2024}.
One of the main ways of manipulating and reading out the states of these spins is electron spin resonance (ESR), where radio-frequency (RF) voltage signals are transduced into oscillating magnetic fields that interact with the spins \cite{lund_principles_2011}.
The response of the spins to these RF excitations is then detected, with the signal generation and detection being done by an ESR spectrometer \cite{poole_electron_1997}.
This work presents an ESR spectrometer that is both inexpensive and flexible, making it possible to precisely tune to spins across a wide frequency range from $\sim2-15$ GHz, either at a single frequency in monochromatic mode, or simultaneously at any two frequencies in that range in two-tone mode.

As currently configured our spectrometer addresses the 2-6~GHz range, needing only to change the RF circulator to swap between 2-4~GHz and 4-6~GHz, but minor adjustments laid out in Appendix \ref{app:diff_freq} can allow it to address the 6-15~GHz range.
It is also possible to address multiple ranges simultaneously by using different components in two-tone mode as described below.
Our spectrometer excites spin samples by coupling a signal generator, able to produce sinusoidal voltages at a variety of frequencies and powers, with a resonator, which converts the voltage into an oscillating magnetic field applied to the spin sample. 
Any signal from the sample itself is detected using the same resonator, then downconverted using an IQ mixer for readout on an oscilloscope. 
Overall control is provided by a field-programmable gate array (FPGA).

In recent years it has become much easier and more affordable to design ESR spectrometers by coupling high-frequency FPGAs with digital-to-analog converters (DACs) for signal generation, and analog-to-digital converters (ADCs) for detection \cite{kaufmann_dac-board_2013,mcpeak_x-band_2019,shi_x-band_2018,yap_ku_2015,stefanazzi_qick_2022}.
Such designs offer many benefits, including very granular control over the frequency and amplitude of the produced signals.
However, their frequency ranges can be somewhat limited by the sampling rate of the electronics, with, for instance, a 5~GSample/s (GSPS) converter needing to work in Nyquist zone 3 to generate a 6~GHz signal,\cite{shannon_communication_1949} potentially resulting in a reduction in power and the presence of unwanted sidebands.
By trading the granular control for a simpler design using affordable off-the-shelf components, we have developed a spectrometer that has only very rudimentary signal shaping capabilities, mostly constrained to square pulses of varying powers, but that can work at much higher frequencies just by swapping out components to match the desired range.
The relative simplicity of the spectrometer also makes it fairly straightforward to configure and extend its functionality, and to understand how each part of the spectrometer works.

We developed this spectrometer in part to allow for both individual and simultaneous control and readout of the spins in a dimer of molecular nanomagnets (MNMs).
MNMs are a class of magnetic materials allowing for chemical engineering of the properties of the quantum spins, with potential quantum computing applications using the molecules as spin qubits \cite{leuenberger_quantum_2001,friedman_single-molecule_2010,chiesa_molecular_2024}.
Performing multi-qubit gates with MNMs will involve simultaneous ESR on multiple spin qubits with frequencies separated by multiple GHz \cite{national_advancing_2023}.
Our lab has recently focused on MNMs engineered to have atomic clock transitions \cite{collett_clock_2019}, which extend the quantum state lifetime $T_2$ by making the spins less sensitive to local field fluctuations.
Dimers of these MNMs can be used to perform quantum gates \cite{collett_constructing_2020}, but only with a spectrometer that can drive each spin simultaneously, and our spectrometer makes this possible.

Below we describe in detail the composition and performance of our spectrometer.
Most of the components are commercially available, and we will only briefly discuss their role; a more general discussion of spectrometer design can be found elsewhere \cite{poole_electron_1997,lund_principles_2011}.
%TODO: Add that the design can be easily altered to suit a variety of different purposes at low cost

\section{Spectrometer Design}
Using the configuration presented in this work, the spectrometer can produce pulses as short as 20~ns and up to 22~dBm (150~mW) of power at the output to the resonator.
It can switch between continuous-wave (CW) and pulsed signals programmatically, though CW mode is mostly useful for doing preliminary characterization of resonator frequencies, as more sensitive and precise CW measurements can be done with a vector network analyzer.
Our main goal in developing this spectrometer is to perform pulsed experiments, and as such only pulsed signals and results are presented here.

The design of this spectrometer can be broken down into three parts: the source and detection electronics, the resonator portion that directly interacts with the spins, and the FPGA controller.

\subsection{Electronics}
\begin{figure}[t!]
	\includegraphics[width=0.40\textwidth]{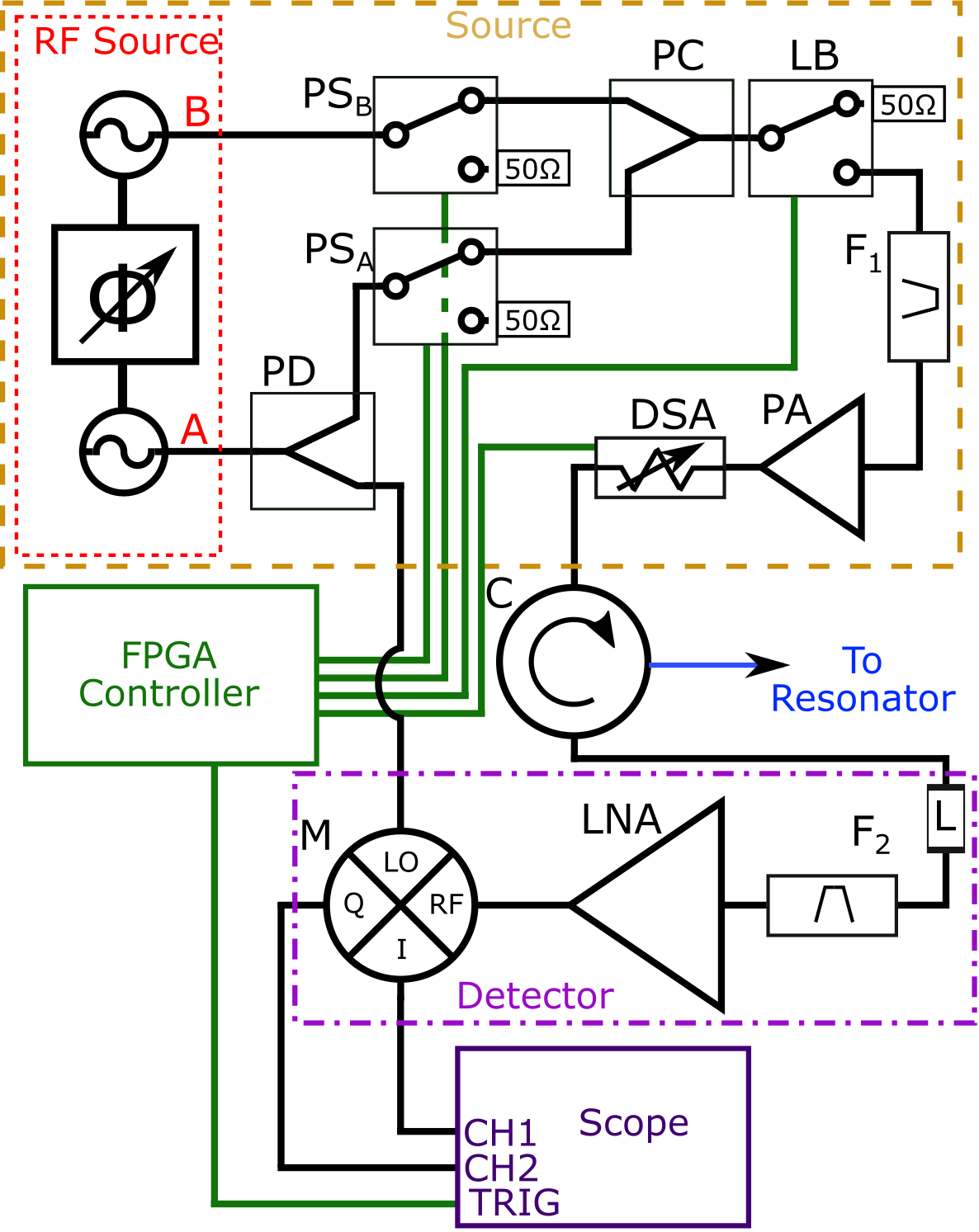}
	\caption{\label{fig:circuit_diagram}A circuit diagram showing the layout and connections of the spectrometer in its monochromatic configuration. Component labels are referenced in the text. See Appendix \ref{app:twotone_elec} for the modifications necessary for two-tone operation.}
\end{figure}

Fig. \ref{fig:circuit_diagram} shows a circuit diagram of the spectrometer electronics. 
Details of the exact off-the-shelf components used can be found in Appendix \ref{app:components}.
%Our RF source (a WindfreakTech SynthHD v1) has two outputs, each of which can output frequencies from 0.054-13.6~GHz, with a range of output powers from below -60~dBm up to 19~dBm.
%our CW sensitivity is fairly poor; performing a full calibration using ESR calibration standards could significantly improve this sensitivity.

\subsubsection{Monochromatic Source}
An RF source generates one or two CW output signals at a single RF frequency (the carrier frequency $\nu_c$), which is our excitation and detection frequency.
One of these signals, which we will label $B$, goes directly into a high-speed RF switch (``pulse switch", PS$_B$ in Fig. \ref{fig:circuit_diagram}), while signal $A$ goes through a power divider (PD), with half of the power going to the detection mixer (see below), and the other half to its own pulse switch (PS$_A$).
Pulsed experiments generally use both outputs to exploit the coherent phase control between the two signals that our source enables, which we will discuss later.
The pulse switches are controlled by the FPGA, opening and closing as needed to generate either CW or pulsed outputs.
The signal from both switches is combined in a power combiner (PC) and optionally put through a bandpass filter (F$_1$) to reduce noise.
We found that even with the pulse switches open there is still enough of the carrier signal leaking through to be detected, so the signal then goes through a blocking switch (LB, for Leakage Block) to more completely attenuate that leaking signal.
This combined signal is then sent through a power amplifier (PA) and a digital step attenuator (DSA).
The DSA is controlled by the FPGA, and allows independent control of the power of each pulse, and potentially shaping the power within a single pulse (that capability is not implemented in the default configuration).
Finally, the signal is routed to an antenna coupled to the resonator through a circulator (C) to prevent reflected power from damaging the amplifier.

The spectrometer can work in reflection mode when a single antenna is coupled to the resonator, and both reflection and transmission mode with two antennas. 
In reflection mode (pictured in Fig. \ref{fig:circuit_diagram}), after the signal has interacted with the resonator it travels back up the same coaxial cable to the circulator, which routes it to the detection electronics. 
In transmission mode, the third output from the circulator is capped with a 50~$\Omega$ terminator, and a coaxial cable attached to the second antenna brings any transmitted signal to the detection electronics.

\subsubsection{Monochromatic Detection}
The returning signal first passes through a power limiter (L) to protect the detection circuitry from the initial high-power pulses.
This protection could also be achieved using another switch, and indeed both configurations have been tested with similar results, but the limiter, being a passive component, is simpler.
We have found that under normal operating conditions there will often not be enough power reaching the detection side to require the limiter, but we leave it in for convenience and safety, as it causes minimal loss ($<0.6$~dB at these frequencies).

The remaining signal optionally passes through another bandpass filter (F$_2$) and then enters a low-noise amplifier (LNA).
At this point, the signal is an RF voltage that has been modified in some way by interactions with the resonator and sample. 
Note that while we output only one frequency, the return signal may have several, due to those interactions. 
It is possible to directly detect signals at these other frequencies by exciting at one frequency and detecting at another, but for this work we will focus on homodyne detection of the signal oscillating at $\nu_c$.
The amplified return signal goes into the RF port of an IQ mixer (M), and with the carrier frequency coming into its LO port the mixer downconverts the signal into low-frequency in-phase (I) and quadrature (Q) signals, which can then be read out on an oscilloscope (Scope).

\subsubsection{Two-tone Electronics}\label{sec:bimod_elec}
We create two-tone pulses with two different output frequencies, $\nu_A$ and $\nu_B$, by simply changing the frequency of one of the RF source outputs.
With the exact same electronics this only provides a two-tone source, as with the above configuration detection is still only possible at $\nu_A$.
To also do two-tone detection, fairly minor changes need to be made to the electronics, adding a splitter and a mixer to output $B$ to match those on output $A$, and then splitting the returning signal after the LNA to send it to the RF ports of both mixers; see Appendix \ref{app:twotone_elec} for a circuit diagram showing these changes.
Note that in this case the spectrometer is doing homodyne detection for both frequencies, though if the frequencies are too close together there will be some crosstalk.

If desired, either or both of the source or detection electronics can be more comprehensively separated to provide more independent two-tone functionality.
For instance, putting in pulses at different frequencies and different powers would require at least separate DSA's, and maximizing the pulse and signal size at each frequency could be done by using two separate antennas both coupled to the resonator, so each frequency gets dedicated power and low-noise amplifiers.
The preliminary two-tone results presented below use the simpler changes described above.

\begin{figure}[t!]
	\includegraphics[width=0.3\textwidth]{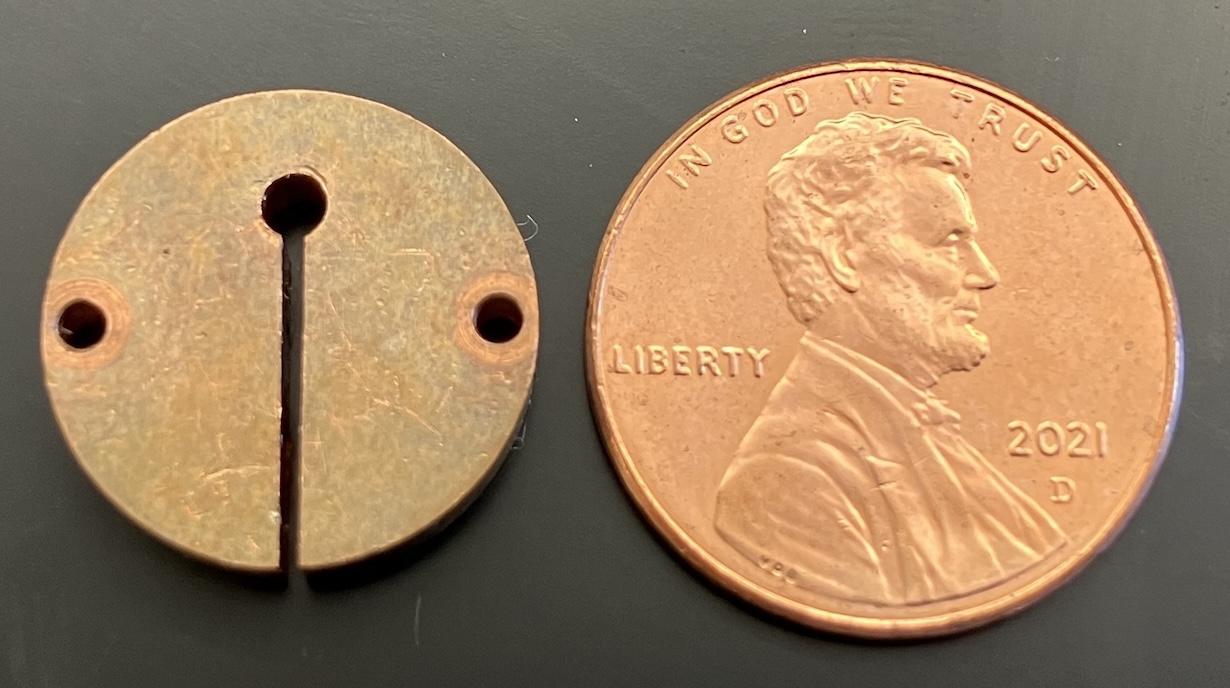}
	\caption{\label{fig:LGR}The loop-gap resonator used for this work, with penny for scale.}
\end{figure}
\subsection{Resonator}
As discussed above, the resonator must be able to couple the spins to an oscillating magnetic field.
For this spectrometer, samples are placed into the loop of a loop-gap resonator (LGR) \cite{froncisz_loop-gap_1982,rinard_loop-gap_2005,eisenach_broadband_2018}.
This style of resonator is formed by shaping a conductor into a connected loop and gap, such that current must flow around the loop to get from one side of the gap to the other. 
The resonator can thus be modeled as a lumped-element RLC circuit with resonant frequency $\nu_r$, with the loop forming the inductor, the gap the capacitor, and the copper body the resistor. 
This design concentrates the oscillating magnetic field inside the loop; the loop dimensions (mm-scale) are generally much smaller than the RF wavelength (cm-scale for GHz frequencies), such that there is a high filling factor for a sample placed in the loop. 
With this design, a relatively low input power translates into a sizable magnetic field strength at the sample, and a small oscillating field from the sample can more easily couple back into the resonator.
In this work, the resonator is a 1/8"-thick disc of OFHC copper 0.6" in diameter with a 0.06" diameter loop and a 0.03" wide gap milled out of it, as shown in Fig. \ref{fig:LGR}.

To interact with the sample, the resonator must transduce an oscillating voltage into an oscillating loop field. 
The voltage is introduced to the resonator via an antenna formed from the exposed inner conductor at the end of a coaxial cable, which is placed near the surface of the resonator. 
A capacitive antenna is made by bending the exposed inner conductor into an `L' shape, then lowering it over the gap, while an inductive antenna sits over the loop and is made by making a circle with the inner conductor and electrically connecting it to the outer conductor. 
The resonator sits inside a copper RF shield at the bottom of a sample probe that is inserted into a cryostat, and coaxial cables transmit signals between the rest of the spectrometer and the resonator through one or more antennas inside the shield. 
For reflection mode a single antenna serves to both excite the resonator and detect signals coming from the sample, and for transmission mode one antenna does the excitation and a different antenna does the detection. 
Transmission mode thus requires two sets of antennas and coaxial cables in the sample probe, which can be terminated if unused. 
A detailed description of a substantially similar antenna-resonator arrangement can be found in Ref. \citenum{joshi_adjustable_2020}.

The resonator is characterized by both $\nu_r$ and the width of its resonance peak $\Delta \nu_r$.
These values combine to give the quality factor of the resonator,
\begin{equation}
	Q=\frac{\nu_r}{\Delta \nu_r}\label{Qnu}
\end{equation}
The resonator used in this work has a room-temperature unloaded $Q$ of 170, and as characterized below a low-temperature $Q$ of $\sim200$.
This low $Q$ is good for pulsed experiments as the pulse shape is less distorted with lower $Q$ values, and also allows us to do two-tone ESR using a monomodal resonator, by operating on either side of the resonance peak, though the frequency separation achieved this way is limited to 10s of MHz.
A truly bimodal resonator\cite{piasecki_bimodal_1996} would enable ESR at frequencies with essentially arbitrary separations, that could be addressed either with one antenna for both frequencies or with dedicated antennas for each.

\subsection{Controller}
Aside from the RF source, most of the components described above are ``dumb'', in that they do not require any control input telling them what to do, they just perform their assigned task with whatever signal inputs they get. 
There are three significant ``smart'' exceptions that take TTL inputs: the switches have control inputs that open and close them; the digital step attenuator has an input that controls the attenuation of the signal; and the oscilloscope has a trigger input. 
All of these control inputs are provided by a low-cost FPGA, running a custom configuration written in the hardware description language Verilog.
For the FPGA's targeted in this work, the control logic can be simulated and verified using open-source tools, enabling fast iteration and development.
As currently configured, the FPGA runs at 200 MHz, setting a minimum timestep of 5~ns for any of the control inputs.
The frequency is configurable, and this frequency was chosen to minimize the timestep while ensuring stable operation of the control program, as running it any faster could create situations where signals propagating within the FPGA would not be able to reach their destinations in time for the next clock tick, introducing random deviations from the programmed behavior.

Using an FPGA in this way provides a great deal of expandability for the spectrometer. 
Adding additional ``smart'' components, or the ability to create more complex pulse sequences, can be done by writing additional control logic into the Verilog configuration. 
Such expansion would likely require the reduction of the clock speed, but increasing the timestep to 10~ns, for example, would allow for a significant amount of expansion, such as the addition of more pulses to perform Carr-Purcell-Meiboom-Gill sequences \cite{meiboom_modified_1958}.

The initial configuration of the FPGA is done by writing out the control logic in Verilog and loading it onto the board. 
All code used to configure and interact with the FPGA is provided in the supplementary materials.
As currently configured, the FPGA supports up to two independent RF output channels, either in CW mode (with one or the other output enabled at any given time), or in pulsed mode with up to three independent pulses on one output channel and up to two independent pulses on the other.
While fully reconfiguring the FPGA takes tens of seconds, a variety of the control parameters are directly adjustable via USB at a fraction of that time.
These parameters will be highlighted when they are introduced in the following section.

\section{Spectrometer Operation}
This section will focus on the important operational details of our spectrometer and will not cover the basics of ESR operation; for a more general overview of both CW and pulsed ESR, see Refs. \citenum{poole_electron_1997} and \citenum{jeschke_instrumentation_2007}.
The general procedure is to find the resonant frequency of the resonator and then perform one or more pulsed experiments to manipulate and probe the spin sample.

Characterization was performed at zero applied magnetic field using the resonator described above with a resonant frequency of $\nu_r=3.902$~GHz, coupled to a capacitive antenna in reflection mode, while immersed in a bath of liquid helium at 3.0~K in a Janis SHI-950T cryostat.
The sample was a solution of the MNM
 Cr$_7$Mn ([Pr$_2$NH$_2$][Cr$_7$MnF$_8$(Piv)$_{15}$(O$_2$C-py]),\cite{larsen_synthesis_2003,timco_heterodimers_2016} diluted to 10\% concentration by volume in toluene, and the input power was 21~dBm.
This sample was chosen based on its zero-field clock transition around 3.9~GHz, our familiarity with it from previous work,\cite{collett_clock_2019} and because future work will explore using dimers of this MNM as spin qubits \cite{collett_constructing_2020}.
All the data presented here was taken on a 4-channel Tektronix MSO24 oscilloscope; monochromatic experiments can be done with a two-channel oscilloscope, but two-tone experiments require either one four-channel scope or two two-channel scopes.
\begin{figure}[b!]
	\includegraphics[width=0.48\textwidth]{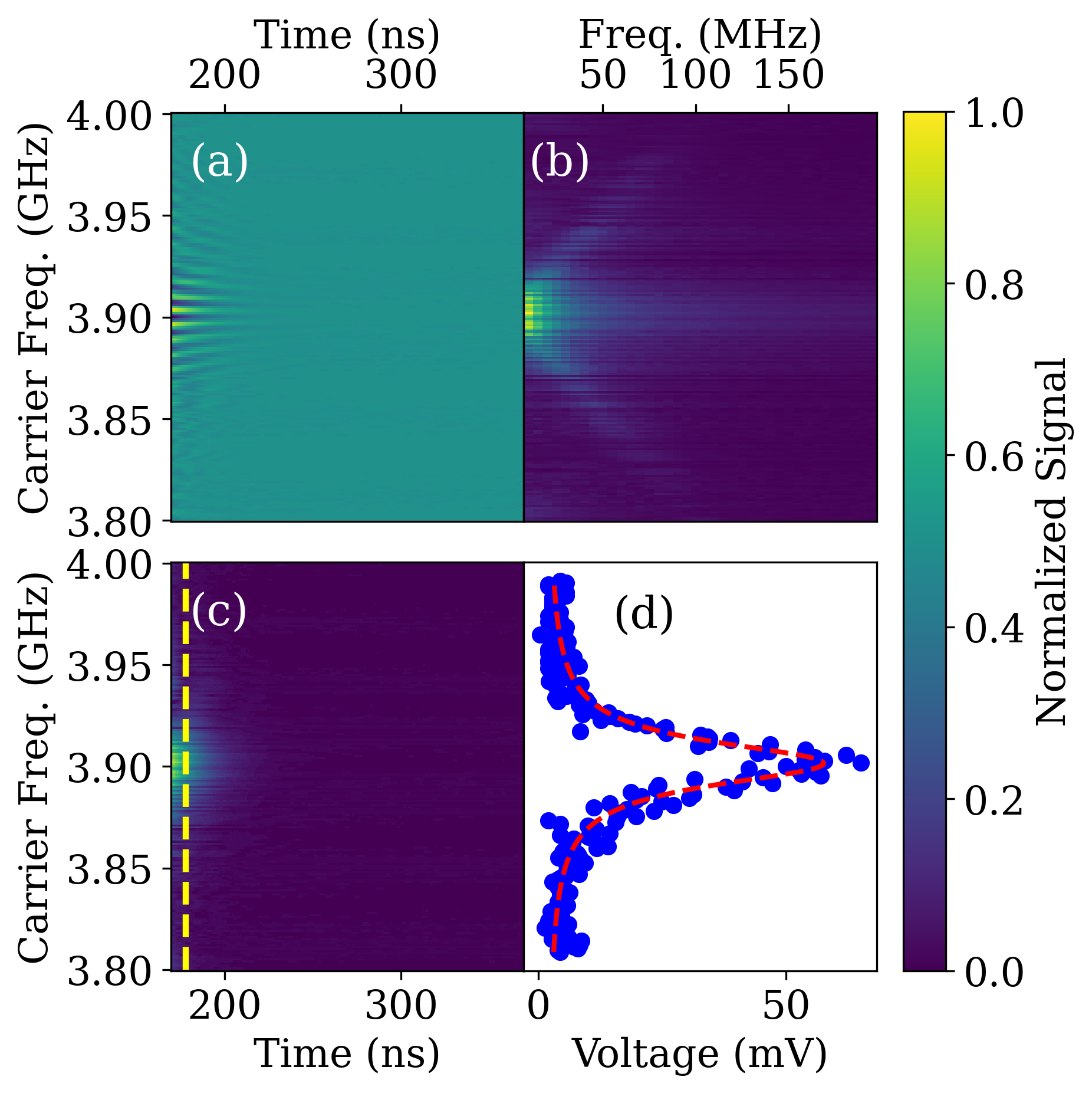}
	\caption{\label{fig:Freq_sweep}(a) Reflected in-phase signal $V_I$ from the resonator showing the ring-down and beating near $\nu_r$. (b) Fourier transform of the signal from (a), showing the weight of different frequencies in the reflected signal shifting as $\nu_c$ is detuned from $\nu_r$. The maximum Fourier weight occurs at the beat frequency $\nu_b=|\nu_c-\nu_r|$. (c) Reflected amplitude signal $V_\text{Amp}$. (d) Vertical cut of amplitude signal from (c) taken at 178~ns, along the dashed yellow line. Red dashed line is a Lorentzian fit, yielding $\nu_r=3.902$~GHz and $\Delta\nu_r=29$~MHz.}
\end{figure}

\subsection{Determining $\nu_r$}
As previously mentioned, the current version of this spectrometer uses CW ESR only for preliminary resonator characterization, letting us scan the input frequency for dips in the background associated with the resonant frequency $\nu_r$.
Precisely determining $\nu_r$ is done by sending in a short ($\sim50$~ns) pulse at $\nu_c$ and monitoring the response immediately after it ends.
When $\nu_c$ is far from $\nu_r$, there will be effectively no signal. 
However, when $\nu_c$ is at or near $\nu_r$, the resonator will absorb energy from the pulse, and when the pulse ends that energy will radiate back into the antenna over a decay time
\begin{equation}
t_\text{decay}=\frac{Q}{\pi\nu_r}\label{Q}
\end{equation}
%where $\Delta \nu_r$ is the width of the resonance peak.
That produces a ``ring-down" signal near $\nu_r$; Fig. \ref{fig:Freq_sweep}(a) shows the in-phase (I) portion of such a signal for the resonator used in this work, and Fig. \ref{fig:Freq_sweep}(c) shows the amplitude of the same signal, calculated from the in-phase ($V_I$) and quadrature voltages ($V_Q$) as
\begin{equation}\label{eq:Vamp}
V_\text{Amp}=\sqrt{V_I^2+V_Q^2}
\end{equation}

Finding $\nu_r$ from this data can be done in two ways.
The first takes a fixed-time cut of the amplitude data and fits that voltage versus the carrier frequency to a Lorentzian function, giving the central frequency $\nu_r$ and the width of the peak $\Delta \nu_r$.
An example of this is shown in Fig. \ref{fig:Freq_sweep}(d), for the vertical cut indicated by the yellow dashed line in (c) at 178~ns.
The second method uses the oscillations that are present in either the I or Q signals, as seen in Fig. \ref{fig:Freq_sweep}(a).
Applying a carrier frequency $\nu_c$ that is detuned from the resonator frequency $\nu_r$ results in a time-dependent phase shift, oscillating the signal between I and Q at the beat frequency $\nu_b=|\nu_c-\nu_r|$.
Taking the Fourier transform of the I or Q data, as shown in Fig. \ref{fig:Freq_sweep}(b), allows the extraction of that beat frequency as a function of $\nu_c$, and fitting that data to find where it goes to zero gives $\nu_r$.
Once $\nu_r$ is found by either method, the quality factor $Q$ can be found by simply fitting the ring-down signal at $\nu_r$ to a decaying exponential and plugging the resulting time constant and frequency into Eq. \ref{Q}.
For the resonator used in this work, the Lorentzian and Fourier methods give resonant frequencies of $\nu_r=3901.6\pm0.3$~MHz and $\nu_r=3901.8\pm0.5$~MHz, respectively, and the quality factor is $Q=198\pm1$.

\subsection{Pulse Generation}
Once $\nu_r$ is determined, a variety of spin echo experiments can be done by generating one or more sets of pulses, depending on whether the spectrometer is operating at a single frequency on one or two outputs (monochromatic), or at two different frequencies for the two outputs (two-tone).

\subsubsection{Monochromatic Pulses}
As noted above, the spectrometer can create up to three pulses at a single frequency, using either one or both outputs; the third pulse is currently only enabled on output $B$.
The default configuration has two of the pulses forming a Hahn echo pulse sequence, the first having a width of $t_{\pi/2}$ and the second a width of $t_{\pi}$, with a delay $\tau$ between them.
The third pulse is optionally used as a nutation pulse, occurring an amount of time $\tau_\text{nut}$ before the first Hahn pulse and having a width of $t_\text{nut}$.
As currently configured, all pulses are programmatically-adjustable in increments of the timestep, 5~ns, with a minimum value of 20~ns and a maximum value of 327.68~$\mu$s, with the exception of $t_\text{nut}$ which has a maximum value of 1.28~µs; the 20~ns minimum pulse time is set by the on/off time of our switches.
The delays between any of the pulses are adjustable from 5~ns to 327.68~$\mu$s, and the sequence repetition time $T$ is adjustable from 5~ns to 21.47~s.
The first Hahn pulse is also attenuated by 3 dB relative to the other pulses, giving it half the power so that the width of the two Hahn pulses can be the same.
The scope trigger rising edge is coincident with the rising edge of the first Hahn pulse; the included acquisition code is configured to trigger the scope on the rising edge and automatically offset to put the expected echo on the scope screen. 

\subsubsection{Two-tone Pulses}
The two RF outputs can also produce two different frequencies, either for two-tone ESR or for heterodyne detection.
As discussed in the two-tone electronics section (\ref{sec:bimod_elec}), minor changes to the electronics are required for simultaneous detection at both frequencies.
The spectrometer can create one or two pulses for each output, and as with the monochromatic setup output $B$ can also have a third pulse.
The details are mostly same as with the monochromatic pulses, and the extra attenuation is timed to coincide with the first Hahn pulse on output $A$.
It is trivial to swap the frequencies of each output.

\begin{figure}[b!]
	\includegraphics[width=0.35\textwidth]{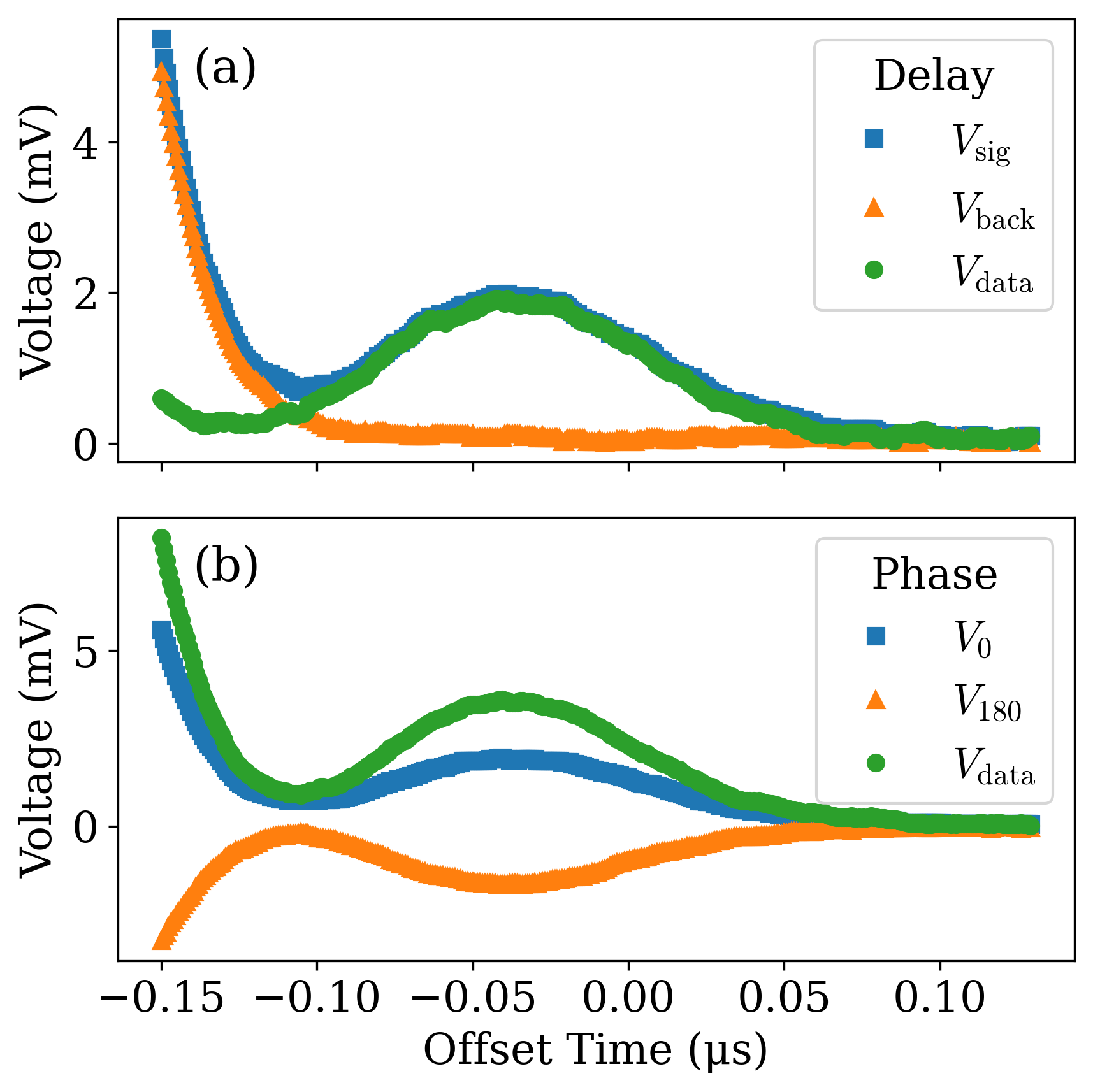}
	\caption{\label{fig:Sub_comp}Comparison of our two methods for background subtraction, based on a Hahn echo sequence with 70~ns pulse widths and inter-pulse delay $\tau=150$~ns to include some of the resonator ring-down. Methods are described in the main text. (a) Delay-based background subtraction, with signal trace $V_\text{sig}$ shown as blue squares, background trace $V_\text{back}$ as orange triangles, and background-subtracted data $V_\text{data}=V_\text{sig}-V_\text{back}$ as green circles. (b) Phase-cycling background subtraction, with $\Delta\phi=0$ ($V_0$) shown as blue squares, $\Delta\phi=180^\circ$ ($V_{180}$ as orange triangles, and background-subtracted data $V_\text{data}=V_\text{0}-V_\text{180}$ as green circles. This removes some portion of the background and doubles the height of the echo.}
\end{figure}
\subsection{Background Subtraction}
Once $\nu_r$ is determined, a Hahn echo sequence is performed.
If there is a strong enough echo signal over the desired experimental parameters, experiments can be done directly by integrating the signal over an appropriate range. 
Unfortunately, for many situations the signal to noise ratio is too low for this to be an effective approach, due to a variety of factors including small samples and large amounts of background noise.
To address these issues we have two main tools: averaging over multiple signals, and background subtraction.
The oscilloscope has built-in averaging, but it is often necessary to take more averages than the maximum provided by the scope.
Our acquisition software in the supplementary materials enables additional averaging as needed, with no upper limit (the functional upper limit is the amount of time required).

If white noise is the only factor making it difficult to see the signal, averaging can suffice to bring out the signal.
However, there can also be systematic offsets, coming from the other electronics for a variety of reasons, and averaging will not remove these.
Instead, we combine averaging with background subtraction, where we take two different measurements where ideally the only thing that changes is the spin echo signal, and then subtract one from the other to remove a variety of non-sample-dependent background effects, such as ground fluctuations when the RF switches open and close, or cavity ring-down.
We do this either by changing the Hahn echo delay $\tau$, or by changing the relative phase between outputs $A$ and $B$.
Not all backgrounds can be removed with these methods, but they have proved quite effective at removing a range of electronic artifacts that would otherwise obscure the spin echo signal; because the resonator ring-down depends on the phase of the input signal, it can only be subtracted using the delay-based method.
For our background subtraction comparison, shown in Fig. \ref{fig:Sub_comp}, we chose $\tau$ such that the background includes some resonator ring-down.
Note that delay-based subtraction can be done with any configuration of the spectrometer, but phase-cycling requires a monochromatic experiment with output $A$ providing the reference frequency (going into the LO port of the mixer) and output $B$ providing the excitation pulses.
Phase-cycling generally results in a higher signal to noise ratio, when the background is not phase-dependent.

For our delay-based background subtraction, we set up our Hahn echo sequence and take a signal measurement $V_\text{sig}$, as shown by the blue squares in Fig. \ref{fig:Sub_comp}(a).
We then change $\tau$ (increasing it by $5$~µs if $\tau<10$~µs, otherwise decreasing it by $5$~µs) but keep the scope window fixed to shift the expected echo signal out of the scope window while leaving the background mostly unchanged, and take a background measurement $V_\text{back}$ as shown by the orange triangles in Fig. \ref{fig:Sub_comp}(a)).
Subtracting $V_\text{back}$ from $V_\text{sig}$ gives us our data, $V_\text{data}=V_\text{sig}-V_\text{back}$, as shown by the green circles in Fig. \ref{fig:Sub_comp}(a), where the resonator ring-down background has been mostly removed.

For our phase-cycled background subtraction, we exploit the fact that if we shift the phase of an input signal by $\Delta\phi=180^\circ$, the portion of the output signal that was downconverted from the carrier frequency (and thus was coming from the resonator and sample) will be inverted.
With our Hahn echo sequence set up and with a relative phase of $\Delta\phi=0$, we can take an in-phase measurement $V_0$, as shown by the blue squares in Fig. \ref{fig:Sub_comp}(b).
If we then change the relative phase to $\Delta\phi=180^\circ$, we can take an inverted measurement $V_{180}$, as shown by the orange triangles in Fig. \ref{fig:Sub_comp}(b).
The low-frequency background in each measurement will be largely identical, because much of it does not depend on the phase, but the echo signal will now be inverted; as noted above, one background that does depend on the phase is resonator ring-down, so if that significantly contributes to the background in an experiment the delay-based background subtraction should be used. 
If we then take $V_\text{data}=V_{180}-V_0$ we will be subtracting the background but doubling the echo signal, as shown by the green circles in Fig. \ref{fig:Sub_comp}(b); this data is usually halved to maintain a constant signal size.
The resonator ring-down background, being phase-dependent, is still present.

\subsection{Standard Experiments}
While many different experiments can be made with the spectrometer as currently configured, here we present several that are included as turn-key options in the acquisition code.
Many other experiments using 1-5 pulses can be set up with minor changes to the acquisition code, and experiments using more pulses can be implemented by changing the FPGA configuration.
For each experiment, phase-cycling was used for background subtraction, with delay times $\tau$ chosen to exclude any resonator ring-down from the background.
\begin{figure}[t!]
	\includegraphics[width=0.45\textwidth]{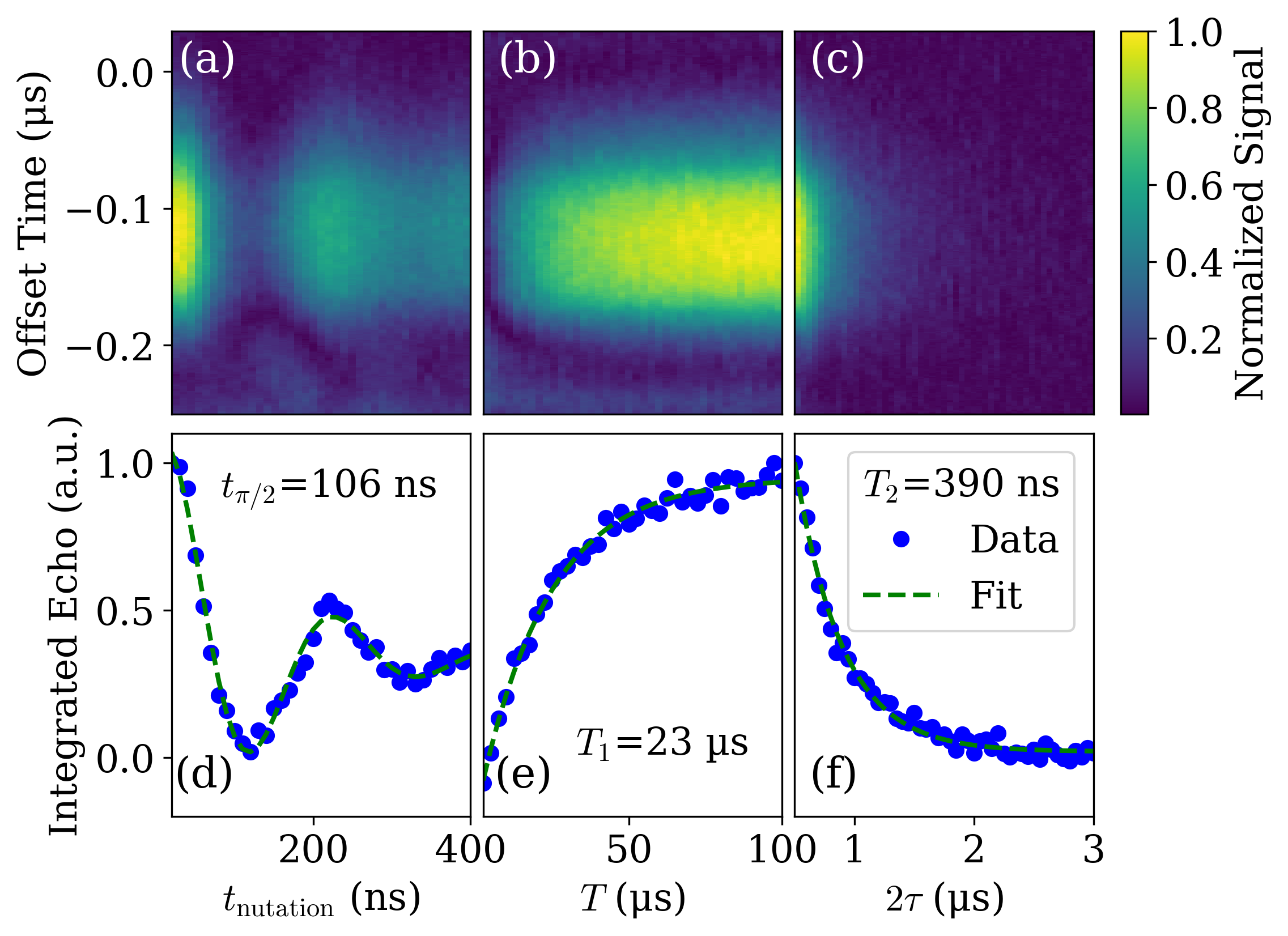}
	\caption{\label{fig:Mono_data}Characterization of monochromatic spin echo experiments. All data is phase-cycled. In all lower panels, integrated echo area is shown as blue circles, while appropriate fits to the echo area are shown as dashed green lines. Data (a) and fitted echo area (d) for a Rabi oscillation experiment, sweeping the nutation pulse width $t_\text{nut}$ to find the appropriate $\pi/2$ pulse width $t_{\pi/2}$. Data (b) and fitted echo area (e) for an inversion recovery experiment, sweeping the delay $T$ between an inverting nutation pulse and the $\pi/2$ pulse to measure $T_1$. Data (c) and fitted echo area (f) for a Hahn echo delay sweep, sweeping the inter-pulse delay $\tau$ to measure $T_2$.}
\end{figure}

\subsubsection{Monochromatic Experiments}
Three different single-frequency experiments are shown in Fig. \ref{fig:Mono_data}.
In addition to these experiments, the acquisition software is also configured to easily do pulse sweep measurements (sweeping $t_{\pi/2}$ and $t_\pi$), phase sweep measurements (sweeping $\Delta\phi$), and frequency sweep measurements (sweeping $\nu_c$).

In (a), we present a Rabi oscillation experiment.
A Hahn echo pulse sequence is set up with $t_{\pi/2}=t_\pi=110$~ns (based on a rough preliminary sweep of the pulse widths) and $\tau=250$~ns.
A nutation pulse is sent 5~µs before the $\pi/2$ pulse, and the width of the nutation pulse $t_\text{nut}$ is swept from 0 to 400~ns.
Integrating the area under the echo results in the data shown in \ref{fig:Mono_data}(d), and fitting that data to a decaying sinusoid (the green dashed line) returns a $\pi/2$ pulse time of $106\pm2$~ns.
\begin{figure}[b!]
	\includegraphics[width=0.48\textwidth]{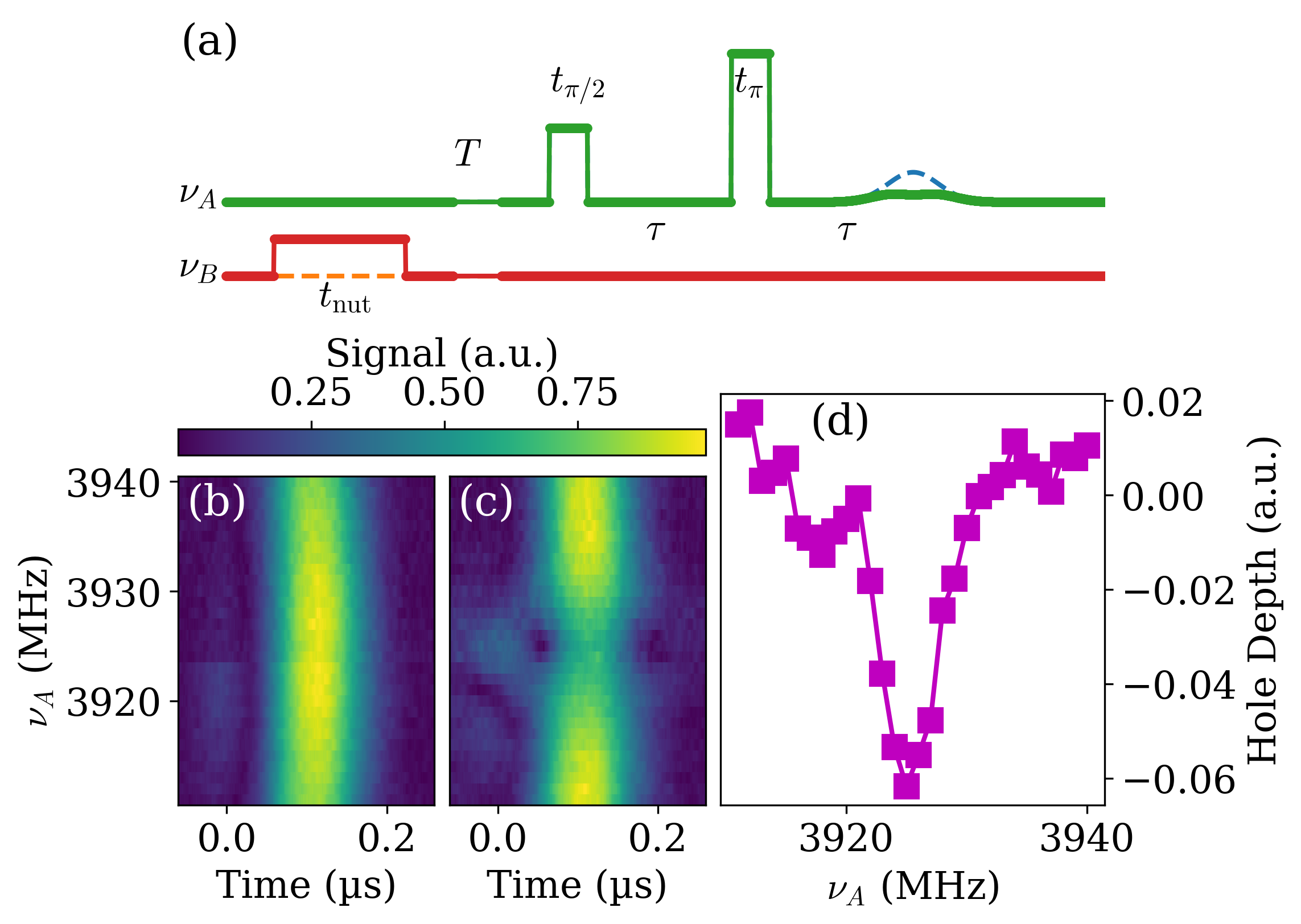}
	\caption{\label{fig:Holeburn}Demonstration of a two-tone hole-burning experiment. (a) Pulse sequence schematic for the experiment, showing the nutation pulse at $\nu_B$ and the Hahn echo sequence at $\nu_A$. The time between the nutation pulse and the first Hahn echo pulse is truncated for clarity. (b,c) Background-subtracted echo data without (b) and with (c) the nutation pulse, using the parameters given in the text. (d) Depth of the hole as a function of frequency. The hole width is $5.6\pm0.7$~MHz.}
\end{figure}

In (b), we present an inversion recovery experiment.
$t_{\pi/2}$ and $t_\pi$ are both set to the 110~ns pulse width determined in the Rabi experiment, with the same $\tau$ as in (a), and a nutation pulse of width $t_\text{nut}=110$~ns; because the attenuator only adds attenuation to the $\pi/2$ pulse, this nutation pulse performs a $\pi$ rotation.
The delay between the nutation pulse and the $\pi/2$ pulse, $T$, is swept from 5-100~µs to determine the longitudinal relaxation rate $T_1$.
Fitting the integrated echo to an exponential (as shown in Fig. \ref{fig:Mono_data}(e)) yields $T_1=23\pm1$~µs.

Finally, in (c) we present a Hahn echo delay sweep.
The same Hahn echo pulse times from (b) are used, but now $\tau$ is swept from 250-1500~ns to determine the transverse relaxation rate $T_2$.
Fig. \ref{fig:Mono_data}(f) shows the integrated echo data and an exponential fit that yields $T_2=390\pm10$~ns (this value is limited by our temperature and level of dilution).

\subsubsection{Two-tone Experiments}
With two separate output frequencies, all of the monochromatic experiments can be done on each frequency individually (with the exception of the phase sweep), but it is also possible to do inherently multi-frequency measurements.
As noted above, only delay-based background subtraction is possible with two-tone experiments.
Our resonator has a low enough $Q$ that we are able to detune more than $40$~MHz from $\nu_r$, and we have tested both double-electron-electron-resonance (DEER) pulses and two-tone hole-burning (HB) experiments.
Our sample, however, is made up of monomers of Cr$_7$Mn, so even though we can excite spins at two frequencies, there is no coupling between those spins and thus there is no DEER signal.
However, we can demonstrate that pulses at both frequencies are working properly with a HB experiment \cite{dzuba_selective_1996}.
As shown in Fig. \ref{fig:Holeburn}(a), a nutation pulse with width $t_\text{nut}$ (so labeled because for this spectrometer it is the same nutation pulse as that used in the Rabi experiment) is sent at one frequency such that it performs a $\pi$ rotation to ``burn a hole'' in the spin spectrum at that frequency, followed after time $T$ by a Hahn echo sequence at the other frequency.
The initial nutation pulse is longer and lower-power than the Hahn pulses, to have a narrower bandwidth.
Integrating the echo signal both with (solid lines in Fig. \ref{fig:Holeburn}(a)) and without (dashed lines) the nutation pulse, and then subtracting one from the other, allows the hole to be characterized, with variation of either frequency, the nutation width and power, or delay, yielding information about the pulse bandwidth and/or spin diffusion in the sample.

For our HB experiment we kept $\nu_B$ constant at 3925~MHz and varied $\nu_A$ from $3911-3940$ MHz, and the results are shown in Fig. \ref{fig:Holeburn}(b-d); note that our echo signal is $\sim23$~MHz above $\nu_r$ due to the interaction between the resonant width of the resonator and the inhomogeneous broadening of our Cr$_7$Mn sample, whose central spin transition frequency is $\sim3.95$~GHz.
Our nutation pulse had a width of $t_\text{nut}=275$~ns and a power of 15 dBm, with $T=5$~$\mu$s, $t_{\pi/2}=t_\pi=80$~ns, and $\tau=300$~ns.
Echo signals, using delay-based background subtraction, are shown both without (b) and with (c) the nutation pulse.
The signal from (b) is subtracted from (c) and the resulting data is integrated to show the hole in Fig. \ref{fig:Holeburn}(d).
A Lorentzian fit to the hole gives a width of $5.6\pm0.7$~MHz, much narrower than the overall echo linewidth of $\sim75$~MHz.
This result demonstrates the ability of the spectrometer to operate at two frequencies simultaneously.

\section{Conclusions}
The spectrometer described herein is effective, flexible, and inexpensive. 
In its current form it can perform many common ESR experiments, and can be relatively easily extended to incorporate additional measurements in the future. 
As of the time of writing, the components of the spectrometer itself (not including the cryostat, oscilloscope, or DC voltage source) cost about \cost, with the majority of that coming from the RF source, the mixer, and the amplifiers.
With the appropriate choice of resonator, it enables experiments both within and outside the standard ESR frequency bands. 
Additionally, the multiple-output nature of the spectrometer can enable two-tone ESR, where two different frequencies are sent to the resonator simultaneously. 
As demonstrated, the spectrometer itself currently supports such experiments, and our lab is working on bimodal resonator designs that will allow simultaneous ESR for spin samples with frequencies separated by more than 1~GHz.

\section{Supplementary Material}
We provide the code used to program the FPGA, as well as the data acquisition code. Instrument drivers are provided for all equipment used, including the temperature controller/thermometer (Lakeshore 335) and programmable power supply (GPD-3303S).

\begin{acknowledgments}
We thank Jonathan Friedman, Brendan Sheehan, Kai-Isaak Ellers, and Andrew Mounce for useful conversations and advice, and Richard Winpenny for samples and useful conversations. Part of this work was enabled by the use of pyscan (github.com/sandialabs/pyscan), scientific measurement software made available by the Center for Integrated Nanotechnologies, an Office of Science User Facility operated for the U.S. Department of Energy. Support for this work was provided by the Hamilton College Dean of Faculty, the Muhlenberg College Provost, and the Muhlenberg College Dean of Students.
\end{acknowledgments}

\section{Author Declarations}
CC: Conceptualization (lead); formal analysis (lead); investigation (lead); methodology (lead); resources (lead); software (lead); supervision (lead); writing – original draft (lead); writing – review and editing (lead). SD: Investigation (supporting); software (supporting); writing – review and editing (supporting). AA: Resources (supporting); writing – review and editing (supporting). GT: Resources (supporting); writing – review and editing (supporting).

The authors have no conflicts to disclose. The data that supports the findings of this study are available from the corresponding author upon reasonable request.

\appendix

\section{\label{app:components}Spectrometer Components}
\begin{longtable}[c]{| p{.1\textwidth} | p{.15\textwidth} | p{.07\textwidth} | p{.07\textwidth} |}% | p{.2\textwidth} |}
	\hline
	Component & Part No. & Cost (\$) & Freq. Range (GHz)\\% & Description \\ [0.5ex] 
	\hline\hline
	RF Source & WindFreakTech SynthHD & 1999 & 0.054-15\\% & Provides two phase-coherent outputs from $\sim-50$ to $\sim19$ dBm\\
	\hline
	RF Switch & MiniCircuits (MC) ZFSWA2R-63DR+ & 88.48 & 0.5-6\\% & Switch for pulse generation, with sub-50~ns switching time\\
	\hline
	RF Splitter & MC ZX10R-14-S+ & 88.48 & 0-10\\% & Power splitter/combiner\\
	\hline
	Power Amplifier & MC ZVA-183-S+ & 1682.17 & 0.7-18\\% & Gives source signal maximum power of ~24 dBm\\
	\hline
	Digital Step Attenuator & MC ZX76-31R75PP-S+ & 166.91 & 0.009-6\\% & Controls source signal power for each pulse separately\\
	\hline
	Circulator & Fairview Microwave (FM) SFC4080B / SFC2040A & 325.99 / 407.99 & 4-8 / 2-4\\% & Circulator chosen based on desired frequency range\\
	\hline
	RF Limiter & MC VLM-63-2W-S+ & 65.71 & 0.03-6\\% & Low loss up to 10 dBm, then significant attenuation\\
	\hline
	Low-Noise Amplifier & FM SLNA-060-40-09-SMA & 1440 & 2-6\\% & 0.9 dB noise floor, 40 dB gain, 12 dBm P1dB\\
	\hline
	IQ Mixer & Marki Microwave MMIQ-0218LXPC & 1925.10 & 2-18\\% & Downconverts returning signal to extract low-frequency modulation caused by the sample\\
	\hline
	FPGA & Lattice ECP5 Evaluation Board & 154.69 & N/A\\
	\hline
	
\end{longtable}

\section{\label{app:diff_freq}Different Frequency Ranges}
Most of the original components work in the 2-6~GHz range, the main exception being the circulator; two circulators are provided above, for the 2-4~GHz range and the 4-8~GHz range. Accessing a higher frequency range (6-15~GHz) can be done by substituting the following components (listed with component name, price, and frequency range):
\begin{itemize}
\item RF switch: Analog Devices HMC547ALP3 (evaluation board EV1HMC547ALP3), \$244.87, DC-20~GHz
\item RF Splitter: This component can be removed, using output A as the mixer LO and output B as the pulsed output.
\item Digital Step Attenuator: This component can be removed, and different tip angles achieved by varying the pulse width.
\item RF Limiter: This component can be removed or replaced with an RF switch.
\item Low-Noise Amplifier: Fairview Microwave FMAM1071, \$2240, 6-18~GHz
\item Circulator: Fairview Microwave SFC0818, \$431.99, 8-18~GHz
\end{itemize}

\section{\label{app:twotone_elec}Two-Tone Electronics Diagram}
\begin{figure}[h!]
	\includegraphics[width=0.40\textwidth]{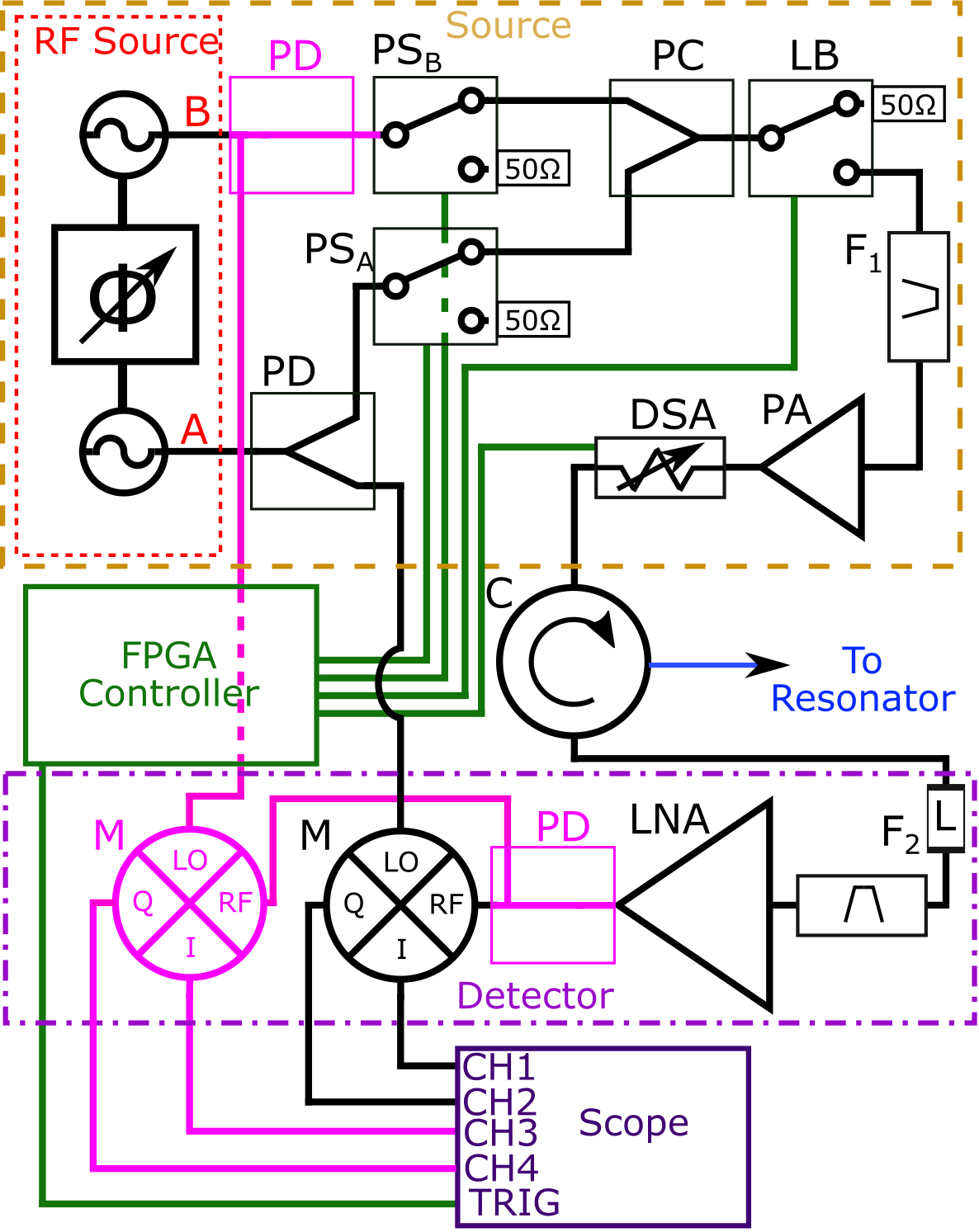}
	\caption{\label{fig:bimod_circuit_diagram}A circuit diagram showing the layout and connections of the spectrometer in its two-tone configuration. Additions for two-tone operation are shown in pink.}
\end{figure}

\bibliography{spectrometer}

%merlin.mbs apsrev4-1.bst 2010-07-25 4.21a (PWD, AO, DPC) hacked
%Control: key (0)
%Control: author (8) initials jnrlst
%Control: editor formatted (1) identically to author
%Control: production of article title (-1) disabled
%Control: page (0) single
%Control: year (1) truncated
%Control: production of eprint (0) enabled
\begin{thebibliography}{26}%
\makeatletter
\providecommand \@ifxundefined [1]{%
 \@ifx{#1\undefined}
}%
\providecommand \@ifnum [1]{%
 \ifnum #1\expandafter \@firstoftwo
 \else \expandafter \@secondoftwo
 \fi
}%
\providecommand \@ifx [1]{%
 \ifx #1\expandafter \@firstoftwo
 \else \expandafter \@secondoftwo
 \fi
}%
\providecommand \natexlab [1]{#1}%
\providecommand \enquote  [1]{``#1''}%
\providecommand \bibnamefont  [1]{#1}%
\providecommand \bibfnamefont [1]{#1}%
\providecommand \citenamefont [1]{#1}%
\providecommand \href@noop [0]{\@secondoftwo}%
\providecommand \href [0]{\begingroup \@sanitize@url \@href}%
\providecommand \@href[1]{\@@startlink{#1}\@@href}%
\providecommand \@@href[1]{\endgroup#1\@@endlink}%
\providecommand \@sanitize@url [0]{\catcode `\\12\catcode `\$12\catcode
  `\&12\catcode `\#12\catcode `\^12\catcode `\_12\catcode `\%12\relax}%
\providecommand \@@startlink[1]{}%
\providecommand \@@endlink[0]{}%
\providecommand \url  [0]{\begingroup\@sanitize@url \@url }%
\providecommand \@url [1]{\endgroup\@href {#1}{\urlprefix }}%
\providecommand \urlprefix  [0]{URL }%
\providecommand \Eprint [0]{\href }%
\providecommand \doibase [0]{http://dx.doi.org/}%
\providecommand \selectlanguage [0]{\@gobble}%
\providecommand \bibinfo  [0]{\@secondoftwo}%
\providecommand \bibfield  [0]{\@secondoftwo}%
\providecommand \translation [1]{[#1]}%
\providecommand \BibitemOpen [0]{}%
\providecommand \bibitemStop [0]{}%
\providecommand \bibitemNoStop [0]{.\EOS\space}%
\providecommand \EOS [0]{\spacefactor3000\relax}%
\providecommand \BibitemShut  [1]{\csname bibitem#1\endcsname}%
\let\auto@bib@innerbib\@empty
%</preamble>
\bibitem [{\citenamefont {Nielsen}\ and\ \citenamefont
  {Chuang}(2010)}]{nielsen_quantum_2010}%
  \BibitemOpen
  \bibfield  {author} {\bibinfo {author} {\bibfnamefont {M.~A.}\ \bibnamefont
  {Nielsen}}\ and\ \bibinfo {author} {\bibfnamefont {I.~L.}\ \bibnamefont
  {Chuang}},\ }\href@noop {} {\emph {\bibinfo {title} {Quantum computation and
  quantum information}}},\ \bibinfo {edition} {10th}\ ed.\ (\bibinfo
  {publisher} {Cambridge University Press},\ \bibinfo {address} {Cambridge ;
  New York},\ \bibinfo {year} {2010})\BibitemShut {NoStop}%
\bibitem [{\citenamefont {Loss}\ and\ \citenamefont
  {DiVincenzo}(1998)}]{loss_quantum_1998}%
  \BibitemOpen
  \bibfield  {author} {\bibinfo {author} {\bibfnamefont {D.}~\bibnamefont
  {Loss}}\ and\ \bibinfo {author} {\bibfnamefont {D.~P.}\ \bibnamefont
  {DiVincenzo}},\ }\href {\doibase 10.1103/PhysRevA.57.120} {\bibfield
  {journal} {\bibinfo  {journal} {Physical Review A}\ }\textbf {\bibinfo
  {volume} {57}},\ \bibinfo {pages} {120} (\bibinfo {year} {1998})}\BibitemShut
  {NoStop}%
\bibitem [{\citenamefont {Chiesa}\ \emph {et~al.}(2024)\citenamefont {Chiesa},
  \citenamefont {Santini}, \citenamefont {Garlatti}, \citenamefont {Luis},\
  and\ \citenamefont {Carretta}}]{chiesa_molecular_2024}%
  \BibitemOpen
  \bibfield  {author} {\bibinfo {author} {\bibfnamefont {A.}~\bibnamefont
  {Chiesa}}, \bibinfo {author} {\bibfnamefont {P.}~\bibnamefont {Santini}},
  \bibinfo {author} {\bibfnamefont {E.}~\bibnamefont {Garlatti}}, \bibinfo
  {author} {\bibfnamefont {F.}~\bibnamefont {Luis}}, \ and\ \bibinfo {author}
  {\bibfnamefont {S.}~\bibnamefont {Carretta}},\ }\href {\doibase
  10.1088/1361-6633/ad1f81} {\bibfield  {journal} {\bibinfo  {journal} {Reports
  on Progress in Physics}\ }\textbf {\bibinfo {volume} {87}},\ \bibinfo {pages}
  {034501} (\bibinfo {year} {2024})}\BibitemShut {NoStop}%
\bibitem [{\citenamefont {Lund}\ \emph {et~al.}(2011)\citenamefont {Lund},
  \citenamefont {Shiotani},\ and\ \citenamefont
  {Shimada}}]{lund_principles_2011}%
  \BibitemOpen
  \bibfield  {author} {\bibinfo {author} {\bibfnamefont {A.}~\bibnamefont
  {Lund}}, \bibinfo {author} {\bibfnamefont {M.}~\bibnamefont {Shiotani}}, \
  and\ \bibinfo {author} {\bibfnamefont {S.}~\bibnamefont {Shimada}},\ }\href
  {http://link.springer.com/10.1007/978-1-4020-5344-3} {\emph {\bibinfo {title}
  {Principles and {Applications} of {ESR} {Spectroscopy}}}}\ (\bibinfo
  {publisher} {Springer Netherlands},\ \bibinfo {address} {Dordrecht},\
  \bibinfo {year} {2011})\BibitemShut {NoStop}%
\bibitem [{\citenamefont {Poole}(1997)}]{poole_electron_1997}%
  \BibitemOpen
  \bibfield  {author} {\bibinfo {author} {\bibfnamefont {C.~P.}\ \bibnamefont
  {Poole}},\ }\href@noop {} {\emph {\bibinfo {title} {Electron {Spin}
  {Resonance}: {A} {Comprehensive} {Treatise} on {Experimental}
  {Techniques}}}},\ \bibinfo {edition} {2nd}\ ed.\ (\bibinfo  {publisher}
  {Wiley},\ \bibinfo {address} {New York},\ \bibinfo {year} {1997})\BibitemShut
  {NoStop}%
\bibitem [{\citenamefont {Kaufmann}\ \emph {et~al.}(2013)\citenamefont
  {Kaufmann}, \citenamefont {Keller}, \citenamefont {Franck}, \citenamefont
  {Barnes}, \citenamefont {Glaser}, \citenamefont {Martinis},\ and\
  \citenamefont {Han}}]{kaufmann_dac-board_2013}%
  \BibitemOpen
  \bibfield  {author} {\bibinfo {author} {\bibfnamefont {T.}~\bibnamefont
  {Kaufmann}}, \bibinfo {author} {\bibfnamefont {T.~J.}\ \bibnamefont
  {Keller}}, \bibinfo {author} {\bibfnamefont {J.~M.}\ \bibnamefont {Franck}},
  \bibinfo {author} {\bibfnamefont {R.~P.}\ \bibnamefont {Barnes}}, \bibinfo
  {author} {\bibfnamefont {S.~J.}\ \bibnamefont {Glaser}}, \bibinfo {author}
  {\bibfnamefont {J.~M.}\ \bibnamefont {Martinis}}, \ and\ \bibinfo {author}
  {\bibfnamefont {S.}~\bibnamefont {Han}},\ }\href {\doibase
  10.1016/j.jmr.2013.07.015} {\bibfield  {journal} {\bibinfo  {journal}
  {Journal of magnetic resonance}\ }\textbf {\bibinfo {volume} {235}},\
  \bibinfo {pages} {10.1016/j.jmr.2013.07.015} (\bibinfo {year}
  {2013})}\BibitemShut {NoStop}%
\bibitem [{\citenamefont {McPeak}\ \emph {et~al.}(2019)\citenamefont {McPeak},
  \citenamefont {Quine}, \citenamefont {Eaton},\ and\ \citenamefont
  {Eaton}}]{mcpeak_x-band_2019}%
  \BibitemOpen
  \bibfield  {author} {\bibinfo {author} {\bibfnamefont {J.~E.}\ \bibnamefont
  {McPeak}}, \bibinfo {author} {\bibfnamefont {R.~W.}\ \bibnamefont {Quine}},
  \bibinfo {author} {\bibfnamefont {S.~S.}\ \bibnamefont {Eaton}}, \ and\
  \bibinfo {author} {\bibfnamefont {G.~R.}\ \bibnamefont {Eaton}},\ }\href
  {\doibase 10.1063/1.5043316} {\bibfield  {journal} {\bibinfo  {journal}
  {Review of Scientific Instruments}\ }\textbf {\bibinfo {volume} {90}},\
  \bibinfo {pages} {024102} (\bibinfo {year} {2019})}\BibitemShut {NoStop}%
\bibitem [{\citenamefont {Shi}\ \emph {et~al.}(2018)\citenamefont {Shi},
  \citenamefont {Mu}, \citenamefont {Qin}, \citenamefont {Dai}, \citenamefont
  {Rong},\ and\ \citenamefont {Du}}]{shi_x-band_2018}%
  \BibitemOpen
  \bibfield  {author} {\bibinfo {author} {\bibfnamefont {Z.}~\bibnamefont
  {Shi}}, \bibinfo {author} {\bibfnamefont {S.}~\bibnamefont {Mu}}, \bibinfo
  {author} {\bibfnamefont {X.}~\bibnamefont {Qin}}, \bibinfo {author}
  {\bibfnamefont {Y.}~\bibnamefont {Dai}}, \bibinfo {author} {\bibfnamefont
  {X.}~\bibnamefont {Rong}}, \ and\ \bibinfo {author} {\bibfnamefont
  {J.}~\bibnamefont {Du}},\ }\href {\doibase 10.1063/1.5048551} {\bibfield
  {journal} {\bibinfo  {journal} {Review of Scientific Instruments}\ }\textbf
  {\bibinfo {volume} {89}},\ \bibinfo {pages} {125104} (\bibinfo {year}
  {2018})}\BibitemShut {NoStop}%
\bibitem [{\citenamefont {Yap}\ \emph {et~al.}(2015)\citenamefont {Yap},
  \citenamefont {Tabuchi}, \citenamefont {Negoro}, \citenamefont {Kagawa},\
  and\ \citenamefont {Kitagawa}}]{yap_ku_2015}%
  \BibitemOpen
  \bibfield  {author} {\bibinfo {author} {\bibfnamefont {Y.~S.}\ \bibnamefont
  {Yap}}, \bibinfo {author} {\bibfnamefont {Y.}~\bibnamefont {Tabuchi}},
  \bibinfo {author} {\bibfnamefont {M.}~\bibnamefont {Negoro}}, \bibinfo
  {author} {\bibfnamefont {A.}~\bibnamefont {Kagawa}}, \ and\ \bibinfo {author}
  {\bibfnamefont {M.}~\bibnamefont {Kitagawa}},\ }\href {\doibase
  10.1063/1.4922791} {\bibfield  {journal} {\bibinfo  {journal} {Review of
  Scientific Instruments}\ }\textbf {\bibinfo {volume} {86}},\ \bibinfo {pages}
  {063110} (\bibinfo {year} {2015})}\BibitemShut {NoStop}%
\bibitem [{\citenamefont {Stefanazzi}\ \emph {et~al.}(2022)\citenamefont
  {Stefanazzi}, \citenamefont {Treptow}, \citenamefont {Wilcer}, \citenamefont
  {Stoughton}, \citenamefont {Bradford}, \citenamefont {Uemura}, \citenamefont
  {Zorzetti}, \citenamefont {Montella}, \citenamefont {Cancelo}, \citenamefont
  {Sussman}, \citenamefont {Houck}, \citenamefont {Saxena}, \citenamefont
  {Arnaldi}, \citenamefont {Agrawal}, \citenamefont {Zhang}, \citenamefont
  {Ding},\ and\ \citenamefont {Schuster}}]{stefanazzi_qick_2022}%
  \BibitemOpen
  \bibfield  {author} {\bibinfo {author} {\bibfnamefont {L.}~\bibnamefont
  {Stefanazzi}}, \bibinfo {author} {\bibfnamefont {K.}~\bibnamefont {Treptow}},
  \bibinfo {author} {\bibfnamefont {N.}~\bibnamefont {Wilcer}}, \bibinfo
  {author} {\bibfnamefont {C.}~\bibnamefont {Stoughton}}, \bibinfo {author}
  {\bibfnamefont {C.}~\bibnamefont {Bradford}}, \bibinfo {author}
  {\bibfnamefont {S.}~\bibnamefont {Uemura}}, \bibinfo {author} {\bibfnamefont
  {S.}~\bibnamefont {Zorzetti}}, \bibinfo {author} {\bibfnamefont
  {S.}~\bibnamefont {Montella}}, \bibinfo {author} {\bibfnamefont
  {G.}~\bibnamefont {Cancelo}}, \bibinfo {author} {\bibfnamefont
  {S.}~\bibnamefont {Sussman}}, \bibinfo {author} {\bibfnamefont
  {A.}~\bibnamefont {Houck}}, \bibinfo {author} {\bibfnamefont
  {S.}~\bibnamefont {Saxena}}, \bibinfo {author} {\bibfnamefont
  {H.}~\bibnamefont {Arnaldi}}, \bibinfo {author} {\bibfnamefont
  {A.}~\bibnamefont {Agrawal}}, \bibinfo {author} {\bibfnamefont
  {H.}~\bibnamefont {Zhang}}, \bibinfo {author} {\bibfnamefont
  {C.}~\bibnamefont {Ding}}, \ and\ \bibinfo {author} {\bibfnamefont {D.~I.}\
  \bibnamefont {Schuster}},\ }\href {\doibase 10.1063/5.0076249} {\bibfield
  {journal} {\bibinfo  {journal} {Review of Scientific Instruments}\ }\textbf
  {\bibinfo {volume} {93}},\ \bibinfo {pages} {044709} (\bibinfo {year}
  {2022})}\BibitemShut {NoStop}%
\bibitem [{\citenamefont {Shannon}(1949)}]{shannon_communication_1949}%
  \BibitemOpen
  \bibfield  {author} {\bibinfo {author} {\bibfnamefont {C.}~\bibnamefont
  {Shannon}},\ }\href {\doibase 10.1109/JRPROC.1949.232969} {\bibfield
  {journal} {\bibinfo  {journal} {Proceedings of the IRE}\ }\textbf {\bibinfo
  {volume} {37}},\ \bibinfo {pages} {10} (\bibinfo {year} {1949})}\BibitemShut
  {NoStop}%
\bibitem [{\citenamefont {Leuenberger}\ and\ \citenamefont
  {Loss}(2001)}]{leuenberger_quantum_2001}%
  \BibitemOpen
  \bibfield  {author} {\bibinfo {author} {\bibfnamefont {M.~N.}\ \bibnamefont
  {Leuenberger}}\ and\ \bibinfo {author} {\bibfnamefont {D.}~\bibnamefont
  {Loss}},\ }\href {\doibase 10.1038/35071024} {\bibfield  {journal} {\bibinfo
  {journal} {Nature}\ }\textbf {\bibinfo {volume} {410}},\ \bibinfo {pages}
  {789} (\bibinfo {year} {2001})}\BibitemShut {NoStop}%
\bibitem [{\citenamefont {Friedman}\ and\ \citenamefont
  {Sarachik}(2010)}]{friedman_single-molecule_2010}%
  \BibitemOpen
  \bibfield  {author} {\bibinfo {author} {\bibfnamefont {J.~R.}\ \bibnamefont
  {Friedman}}\ and\ \bibinfo {author} {\bibfnamefont {M.~P.}\ \bibnamefont
  {Sarachik}},\ }\href {\doibase 10.1146/annurev-conmatphys-070909-104053}
  {\bibfield  {journal} {\bibinfo  {journal} {Annual Review of Condensed Matter
  Physics}\ }\textbf {\bibinfo {volume} {1}},\ \bibinfo {pages} {109} (\bibinfo
  {year} {2010})}\BibitemShut {NoStop}%
\bibitem [{\citenamefont {{National Academies of Sciences, Engineering, and
  Medicine}}(2023)}]{national_advancing_2023}%
  \BibitemOpen
  \bibfield  {author} {\bibinfo {author} {\bibnamefont {{National Academies of
  Sciences, Engineering, and Medicine}}},\ }\href
  {https://www.nap.edu/catalog/26850} {\emph {\bibinfo {title} {Advancing
  {Chemistry} and {Quantum} {Information} {Science}: {An} {Assessment} of
  {Research} {Opportunities} at the {Interface} of {Chemistry} and {Quantum}
  {Information} {Science} in the {United} {States}}}}\ (\bibinfo  {publisher}
  {National Academies Press},\ \bibinfo {address} {Washington, D.C.},\ \bibinfo
  {year} {2023})\ \bibinfo {note} {https://doi.org/10.17226/26850}\BibitemShut
  {NoStop}%
\bibitem [{\citenamefont {Collett}\ \emph {et~al.}(2019)\citenamefont
  {Collett}, \citenamefont {Ellers}, \citenamefont {Russo}, \citenamefont
  {Kittilstved}, \citenamefont {Timco}, \citenamefont {Winpenny},\ and\
  \citenamefont {Friedman}}]{collett_clock_2019}%
  \BibitemOpen
  \bibfield  {author} {\bibinfo {author} {\bibfnamefont {C.~A.}\ \bibnamefont
  {Collett}}, \bibinfo {author} {\bibfnamefont {K.-I.}\ \bibnamefont {Ellers}},
  \bibinfo {author} {\bibfnamefont {N.}~\bibnamefont {Russo}}, \bibinfo
  {author} {\bibfnamefont {K.~R.}\ \bibnamefont {Kittilstved}}, \bibinfo
  {author} {\bibfnamefont {G.~A.}\ \bibnamefont {Timco}}, \bibinfo {author}
  {\bibfnamefont {R.~E.~P.}\ \bibnamefont {Winpenny}}, \ and\ \bibinfo {author}
  {\bibfnamefont {J.~R.}\ \bibnamefont {Friedman}},\ }\href {\doibase
  10.3390/magnetochemistry5010004} {\bibfield  {journal} {\bibinfo  {journal}
  {Magnetochemistry}\ }\textbf {\bibinfo {volume} {5}},\ \bibinfo {pages} {4}
  (\bibinfo {year} {2019})}\BibitemShut {NoStop}%
\bibitem [{\citenamefont {Collett}\ \emph {et~al.}(2020)\citenamefont
  {Collett}, \citenamefont {Santini}, \citenamefont {Carretta},\ and\
  \citenamefont {Friedman}}]{collett_constructing_2020}%
  \BibitemOpen
  \bibfield  {author} {\bibinfo {author} {\bibfnamefont {C.~A.}\ \bibnamefont
  {Collett}}, \bibinfo {author} {\bibfnamefont {P.}~\bibnamefont {Santini}},
  \bibinfo {author} {\bibfnamefont {S.}~\bibnamefont {Carretta}}, \ and\
  \bibinfo {author} {\bibfnamefont {J.~R.}\ \bibnamefont {Friedman}},\ }\href
  {\doibase 10.1103/PhysRevResearch.2.032037} {\bibfield  {journal} {\bibinfo
  {journal} {Physical Review Research}\ }\textbf {\bibinfo {volume} {2}},\
  \bibinfo {pages} {032037} (\bibinfo {year} {2020})}\BibitemShut {NoStop}%
\bibitem [{\citenamefont {Froncisz}\ and\ \citenamefont
  {Hyde}(1982)}]{froncisz_loop-gap_1982}%
  \BibitemOpen
  \bibfield  {author} {\bibinfo {author} {\bibfnamefont {W.}~\bibnamefont
  {Froncisz}}\ and\ \bibinfo {author} {\bibfnamefont {J.~S.}\ \bibnamefont
  {Hyde}},\ }\href {\doibase 10.1016/0022-2364(82)90221-9} {\bibfield
  {journal} {\bibinfo  {journal} {Journal of Magnetic Resonance (1969)}\
  }\textbf {\bibinfo {volume} {47}},\ \bibinfo {pages} {515} (\bibinfo {year}
  {1982})}\BibitemShut {NoStop}%
\bibitem [{\citenamefont {Rinard}\ and\ \citenamefont
  {Eaton}(2005)}]{rinard_loop-gap_2005}%
  \BibitemOpen
  \bibfield  {author} {\bibinfo {author} {\bibfnamefont {G.~A.}\ \bibnamefont
  {Rinard}}\ and\ \bibinfo {author} {\bibfnamefont {G.~R.}\ \bibnamefont
  {Eaton}},\ }in\ \href {\doibase 10.1007/0-306-48533-8_2} {\emph {\bibinfo
  {booktitle} {Biomedical {EPR}, {Part} {B}: {Methodology}, {Instrumentation},
  and {Dynamics}}}},\ \bibinfo {series and number} {\bibinfo {series}
  {Biological {Magnetic} {Resonance}}\ No.\ \bibinfo {number} {24/B}},\
  \bibinfo {editor} {edited by\ \bibinfo {editor} {\bibfnamefont {S.~R.}\
  \bibnamefont {Eaton}}, \bibinfo {editor} {\bibfnamefont {G.~R.}\ \bibnamefont
  {Eaton}}, \ and\ \bibinfo {editor} {\bibfnamefont {L.~J.}\ \bibnamefont
  {Berliner}}}\ (\bibinfo  {publisher} {Kluwer Academic/Plenum Publishers},\
  \bibinfo {address} {New York},\ \bibinfo {year} {2005})\ pp.\ \bibinfo
  {pages} {19--52}\BibitemShut {NoStop}%
\bibitem [{\citenamefont {Eisenach}\ \emph {et~al.}(2018)\citenamefont
  {Eisenach}, \citenamefont {Barry}, \citenamefont {Pham}, \citenamefont
  {Rojas}, \citenamefont {Englund},\ and\ \citenamefont
  {Braje}}]{eisenach_broadband_2018}%
  \BibitemOpen
  \bibfield  {author} {\bibinfo {author} {\bibfnamefont {E.~R.}\ \bibnamefont
  {Eisenach}}, \bibinfo {author} {\bibfnamefont {J.~F.}\ \bibnamefont {Barry}},
  \bibinfo {author} {\bibfnamefont {L.~M.}\ \bibnamefont {Pham}}, \bibinfo
  {author} {\bibfnamefont {R.~G.}\ \bibnamefont {Rojas}}, \bibinfo {author}
  {\bibfnamefont {D.~R.}\ \bibnamefont {Englund}}, \ and\ \bibinfo {author}
  {\bibfnamefont {D.~A.}\ \bibnamefont {Braje}},\ }\href {\doibase
  10.1063/1.5037465} {\bibfield  {journal} {\bibinfo  {journal} {Review of
  Scientific Instruments}\ }\textbf {\bibinfo {volume} {89}},\ \bibinfo {pages}
  {094705} (\bibinfo {year} {2018})}\BibitemShut {NoStop}%
\bibitem [{\citenamefont {Joshi}\ \emph {et~al.}(2020)\citenamefont {Joshi},
  \citenamefont {Kubasek}, \citenamefont {Nikolov}, \citenamefont {Sheehan},
  \citenamefont {Costa}, \citenamefont {Allão~Cassaro},\ and\ \citenamefont
  {Friedman}}]{joshi_adjustable_2020}%
  \BibitemOpen
  \bibfield  {author} {\bibinfo {author} {\bibfnamefont {G.}~\bibnamefont
  {Joshi}}, \bibinfo {author} {\bibfnamefont {J.}~\bibnamefont {Kubasek}},
  \bibinfo {author} {\bibfnamefont {I.}~\bibnamefont {Nikolov}}, \bibinfo
  {author} {\bibfnamefont {B.}~\bibnamefont {Sheehan}}, \bibinfo {author}
  {\bibfnamefont {T.~A.}\ \bibnamefont {Costa}}, \bibinfo {author}
  {\bibfnamefont {R.~A.}\ \bibnamefont {Allão~Cassaro}}, \ and\ \bibinfo
  {author} {\bibfnamefont {J.~R.}\ \bibnamefont {Friedman}},\ }\href {\doibase
  10.1063/1.5133074} {\bibfield  {journal} {\bibinfo  {journal} {Review of
  Scientific Instruments}\ }\textbf {\bibinfo {volume} {91}},\ \bibinfo {pages}
  {023104} (\bibinfo {year} {2020})}\BibitemShut {NoStop}%
\bibitem [{\citenamefont {Piasecki}\ \emph {et~al.}(1996)\citenamefont
  {Piasecki}, \citenamefont {Froncisz},\ and\ \citenamefont
  {Hyde}}]{piasecki_bimodal_1996}%
  \BibitemOpen
  \bibfield  {author} {\bibinfo {author} {\bibfnamefont {W.}~\bibnamefont
  {Piasecki}}, \bibinfo {author} {\bibfnamefont {W.}~\bibnamefont {Froncisz}},
  \ and\ \bibinfo {author} {\bibfnamefont {J.~S.}\ \bibnamefont {Hyde}},\
  }\href {\doibase 10.1063/1.1147001} {\bibfield  {journal} {\bibinfo
  {journal} {Review of Scientific Instruments}\ }\textbf {\bibinfo {volume}
  {67}},\ \bibinfo {pages} {1896} (\bibinfo {year} {1996})}\BibitemShut
  {NoStop}%
\bibitem [{\citenamefont {Meiboom}\ and\ \citenamefont
  {Gill}(1958)}]{meiboom_modified_1958}%
  \BibitemOpen
  \bibfield  {author} {\bibinfo {author} {\bibfnamefont {S.}~\bibnamefont
  {Meiboom}}\ and\ \bibinfo {author} {\bibfnamefont {D.}~\bibnamefont {Gill}},\
  }\href {\doibase 10.1063/1.1716296} {\bibfield  {journal} {\bibinfo
  {journal} {Review of Scientific Instruments}\ }\textbf {\bibinfo {volume}
  {29}},\ \bibinfo {pages} {688} (\bibinfo {year} {1958})}\BibitemShut
  {NoStop}%
\bibitem [{\citenamefont {Jeschke}(2007)}]{jeschke_instrumentation_2007}%
  \BibitemOpen
  \bibfield  {author} {\bibinfo {author} {\bibfnamefont {G.}~\bibnamefont
  {Jeschke}},\ }in\ \href {\doibase 10.1007/978-0-387-49367-1_2} {\emph
  {\bibinfo {booktitle} {{ESR} {Spectroscopy} in {Membrane} {Biophysics}}}},\
  \bibinfo {series and number} {Biological {Magnetic} {Resonance}},\ \bibinfo
  {editor} {edited by\ \bibinfo {editor} {\bibfnamefont {M.~A.}\ \bibnamefont
  {Hemminga}}\ and\ \bibinfo {editor} {\bibfnamefont {L.~J.}\ \bibnamefont
  {Berliner}}}\ (\bibinfo  {publisher} {Springer US},\ \bibinfo {address}
  {Boston, MA},\ \bibinfo {year} {2007})\ pp.\ \bibinfo {pages}
  {17--47}\BibitemShut {NoStop}%
\bibitem [{\citenamefont {Larsen}\ \emph {et~al.}(2003)\citenamefont {Larsen},
  \citenamefont {McInnes}, \citenamefont {El~Mkami}, \citenamefont {Overgaard},
  \citenamefont {Piligkos}, \citenamefont {Rajaraman}, \citenamefont
  {Rentschler}, \citenamefont {Smith}, \citenamefont {Smith}, \citenamefont
  {Boote}, \citenamefont {Jennings}, \citenamefont {Timco},\ and\ \citenamefont
  {Winpenny}}]{larsen_synthesis_2003}%
  \BibitemOpen
  \bibfield  {author} {\bibinfo {author} {\bibfnamefont {F.~K.}\ \bibnamefont
  {Larsen}}, \bibinfo {author} {\bibfnamefont {E.~J.~L.}\ \bibnamefont
  {McInnes}}, \bibinfo {author} {\bibfnamefont {H.}~\bibnamefont {El~Mkami}},
  \bibinfo {author} {\bibfnamefont {J.}~\bibnamefont {Overgaard}}, \bibinfo
  {author} {\bibfnamefont {S.}~\bibnamefont {Piligkos}}, \bibinfo {author}
  {\bibfnamefont {G.}~\bibnamefont {Rajaraman}}, \bibinfo {author}
  {\bibfnamefont {E.}~\bibnamefont {Rentschler}}, \bibinfo {author}
  {\bibfnamefont {A.~A.}\ \bibnamefont {Smith}}, \bibinfo {author}
  {\bibfnamefont {G.~M.}\ \bibnamefont {Smith}}, \bibinfo {author}
  {\bibfnamefont {V.}~\bibnamefont {Boote}}, \bibinfo {author} {\bibfnamefont
  {M.}~\bibnamefont {Jennings}}, \bibinfo {author} {\bibfnamefont {G.~A.}\
  \bibnamefont {Timco}}, \ and\ \bibinfo {author} {\bibfnamefont {R.~E.~P.}\
  \bibnamefont {Winpenny}},\ }\href {\doibase 10.1002/anie.200390034}
  {\bibfield  {journal} {\bibinfo  {journal} {Angewandte Chemie International
  Edition}\ }\textbf {\bibinfo {volume} {42}},\ \bibinfo {pages} {101}
  (\bibinfo {year} {2003})}\BibitemShut {NoStop}%
\bibitem [{\citenamefont {Timco}\ \emph {et~al.}(2016)\citenamefont {Timco},
  \citenamefont {Marocchi}, \citenamefont {Garlatti}, \citenamefont {Barker},
  \citenamefont {Albring}, \citenamefont {Bellini}, \citenamefont {Manghi},
  \citenamefont {McInnes}, \citenamefont {Pritchard}, \citenamefont {Tuna},
  \citenamefont {Wernsdorfer}, \citenamefont {Lorusso}, \citenamefont
  {Amoretti}, \citenamefont {Carretta}, \citenamefont {Affronte},\ and\
  \citenamefont {Winpenny}}]{timco_heterodimers_2016}%
  \BibitemOpen
  \bibfield  {author} {\bibinfo {author} {\bibfnamefont {G.}~\bibnamefont
  {Timco}}, \bibinfo {author} {\bibfnamefont {S.}~\bibnamefont {Marocchi}},
  \bibinfo {author} {\bibfnamefont {E.}~\bibnamefont {Garlatti}}, \bibinfo
  {author} {\bibfnamefont {C.}~\bibnamefont {Barker}}, \bibinfo {author}
  {\bibfnamefont {M.}~\bibnamefont {Albring}}, \bibinfo {author} {\bibfnamefont
  {V.}~\bibnamefont {Bellini}}, \bibinfo {author} {\bibfnamefont
  {F.}~\bibnamefont {Manghi}}, \bibinfo {author} {\bibfnamefont {E.~J.~L.}\
  \bibnamefont {McInnes}}, \bibinfo {author} {\bibfnamefont {R.~G.}\
  \bibnamefont {Pritchard}}, \bibinfo {author} {\bibfnamefont {F.}~\bibnamefont
  {Tuna}}, \bibinfo {author} {\bibfnamefont {W.}~\bibnamefont {Wernsdorfer}},
  \bibinfo {author} {\bibfnamefont {G.}~\bibnamefont {Lorusso}}, \bibinfo
  {author} {\bibfnamefont {G.}~\bibnamefont {Amoretti}}, \bibinfo {author}
  {\bibfnamefont {S.}~\bibnamefont {Carretta}}, \bibinfo {author}
  {\bibfnamefont {M.}~\bibnamefont {Affronte}}, \ and\ \bibinfo {author}
  {\bibfnamefont {R.~E.~P.}\ \bibnamefont {Winpenny}},\ }\href {\doibase
  10.1039/C6DT01941B} {\bibfield  {journal} {\bibinfo  {journal} {Dalton
  Transactions}\ }\textbf {\bibinfo {volume} {45}},\ \bibinfo {pages} {16610}
  (\bibinfo {year} {2016})}\BibitemShut {NoStop}%
\bibitem [{\citenamefont {Dzuba}\ and\ \citenamefont
  {Kawamori}(1996)}]{dzuba_selective_1996}%
  \BibitemOpen
  \bibfield  {author} {\bibinfo {author} {\bibfnamefont {S.~A.}\ \bibnamefont
  {Dzuba}}\ and\ \bibinfo {author} {\bibfnamefont {A.}~\bibnamefont
  {Kawamori}},\ }\href {\doibase
  10.1002/(SICI)1099-0534(1996)8:1<49::AID-CMR4>3.0.CO;2-P} {\bibfield
  {journal} {\bibinfo  {journal} {Concepts in Magnetic Resonance}\ }\textbf
  {\bibinfo {volume} {8}},\ \bibinfo {pages} {49} (\bibinfo {year}
  {1996})}\BibitemShut {NoStop}%
\end{thebibliography}%

\end{document}